\documentclass[twocolumn,showpacs,preprintnumbers,amsmath,amssymb]{revtex4}

\usepackage{graphicx}
\usepackage{graphicx,epstopdf}
\usepackage{dcolumn}
\usepackage{bm}

\begin{document}


\title{Basin stability measure of different steady states in coupled oscillators}
\author{Sarbendu Rakshit$^1$}
 \author{Bidesh K. Bera$^1$}
\author{Soumen Majhi$^1$}
\author{Chittaranjan Hens$^2$}
\author{Dibakar Ghosh$^1$}
\email{diba.ghosh@gmail.com}
\affiliation{$^1$Physics and Applied Mathematics Unit, Indian Statistical Institute, 203 B. T. Road, Kolkata-700108, India\\
$^2$Department of Mathematics, Bar-Ilan University, Ramat Gan 52900, Israel}

\date{\today}

\begin{abstract} In this report, we investigate the stabilization of saddle fixed points in coupled oscillators where individual oscillators exhibit the saddle fixed points. The coupled oscillators may have two structurally different types of suppressed states, namely amplitude death and oscillation death. The stabilization of saddle equilibrium point refers to the amplitude death state where oscillations are ceased and all the oscillators converge to the single stable steady state via inverse pitchfork bifurcation. Due to multistability features of oscillation death states, linear stability theory fails to analyze the stability of such states analytically, so we quantify all the states by basin stability measurement which is an universal nonlocal nonlinear concept and it interplays with the volume of basins of attractions. We also observe multi-clustered oscillation death states in a random network and measure them using basin stability framework. To explore such phenomena we choose a network of coupled Duffing-Holmes and Lorenz oscillators which are interacting through mean-field coupling. We investigate how basin stability for different steady states depends on mean-field density and coupling strength.  We also analytically derive stability conditions for different steady states and confirm by rigorous bifurcation analysis.        
\end{abstract}


\maketitle


\section*{Introduction}
Different types of collective behavior emerge when two or more dynamical units interact with each other and suppression of oscillation is one of the most interesting phenomena among them. Oscillation quenched states are categorized in two processes named as amplitude death (AD) \cite{ad_phyrep} and oscillation death (OD) \cite{od_phyrep}. AD state is a result of stable homogeneous steady state (HSS), where all the oscillators merge or converge in one common steady state. In the case of OD state, oscillators populate to different stable steady states which are coupling dependent fixed points termed as stable inhomogeneous steady states (IHSS) and these states are the results of symmetry-breaking bifurcations in coupled oscillators. Also network of coupled oscillators exhibit multi-cluster oscillation death (MCOD)  in nonlocally coupled oscillators \cite{mcod1}. MCOD pattern refers to the stabilization of various coupling dependent steady states to which the oscillators converge. Depending upon the initial conditions of each oscillator, the positions of the stable steady states for MCOD state may vary.  AD state has a great importance to suppress unwanted oscillations.  Such oscillations are responsible for obstructing certain process in some biological systems \cite{neuronal1} and laser experiments \cite{laser1}. Due to ushering of inhomogeneity in homogeneous systems, OD state is very complicated phenomena and closely related to many biological processes  such as cellular differentiation \cite{celular_jtb}, also in neural networks \cite{nural1_PhycD} and synthetic genetic oscillators \cite{syn_gen_prl,syn_gen_epl}. Recently, the transition from AD to OD state via Turing type bifurcations has been articulated \cite{ad-od_prl}. Later many researchers have explored such transition using different types of coupling strategies such as mean-field \cite{ad-od_mfield}, presence of direct and indirect coupling \cite{ad-od_indrct}, mean repulsive interaction \cite{ad-od_repul1}. Also cyclic type of interaction \cite{ad-od_cyclic} can induce AD-OD transition in mismatched coupled systems. Beside IHSS (i.e. OD) state, there are many stable steady states which are also coupling dependent states known as non-trivial homogeneous steady states (NHSS). In ref. \cite{mixed_mode}, the authors discussed about the suppression of mixed mode oscillations state in coupled oscillators. As AD state is a result of stabilization of HSS so it may be easy to derive the analytical condition for stability but in case of coupling dependent stable steady states (OD and NHSS), it is not always possible to obtain the stability condition analytically since OD states are multi-stable by nature.  Most of the previous results on OD states are characterized by only bifurcation analysis and there is no clear discussion about the basin of multi stable OD states. So it is interesting to study the variations of such multi stable steady states with respect to the basins of attractions because multi stable steady states are omnipresent in many coupled dynamical systems. 
\par Up to now, the stability of such collective steady states (AD or OD) in coupled network are characterized by the sign of real parts of eigenvalues of the corresponding Jacobian matrix. This linear stability analysis is valid only for infinitesimal perturbation near the steady states.
So, the linear stability analysis is necessary for the stability of steady state but not sufficient against some significant perturbations. Since non-small perturbation is ubiquitous in nature and many man-made systems, so we need a global measure to characterize the stability. In this context a pioneer work \cite{BS-nat_phys}, they have developed a universal measure in complex systems as {\it basin stability} (BS) which is related to the volume of basin of attraction. The concept of BS has a lot of applications in real-world systems such as power grids \cite{pwr_grd}, arrays of coupled lasers\cite{laser_opt} etc. and effectively applied in many field of science \cite{bs_njp,bs_epjst,bs_ndes} that interplays with the systems which exhibit multi-stability. In practical situation such as human brain \cite{pnas,nat_rev_neu}, cell regulatory network \cite{cell_regu_BD} and many other natural phenomena \cite{pnas_climate,amazon1,amazon2} show the multi-stable behavior \cite{nat_multistable} and also in the economics and social sciences \cite{soc1,soc2,soc3,soc4}, the path dependence processes are suitably described by multistability. To quantify the stability of such multistable states in dynamical systems, the BS measure is successfully applied in finite \cite{BS-nat_phys,bs_finite,bs_finite2} as well as infinite dimensional systems \cite{bs_time_delay}. The BS approach is well studied in various types of emergent and collective behavior in network of dynamical systems such as synchronizability \cite{annul_rev_con} of static and time varying complex network \cite{time_varing} and many others but BS measure in quantification of different multi-stable steady states in coupled systems has not been explored yet, to the best of our knowledge. Therefore, systematic studies on such unnoticed phenomena deserve special attention.
\par In this work, we are dealing with finite dimensional systems and trying to give BS measure for oscillation suppression states (such as AD, OD, NHSS and MCOD) in a network of coupled dynamical systems. Oscillation cessations are significantly applied in many biological and physical processes where unwanted oscillations may arise so we need to suppress the oscillations to some desired stable steady states. We consider a network of globally and randomly connected oscillators through mean-field coupling. This mean field coupling is a natural coupling scheme which is extensively studied for different consequence in physics \cite{phy_meanfield},  ecology \cite{eco_meanfield}, biology \cite{neuronal1,pnas_cell_commun}, chemistry, electrical circuits \cite{ad_phyrep,od_phyrep}. Also this type of interaction arises in metapopulation ecology where by proper tuning of mean-field density parameter, two-patch ecosystems are evolving from an open patchy ecosystems to closed patchy ecosystems \cite{eco_meanfield}. The role of mean-field density is also discussed in Ref. \cite{pnas_cell_commun,syn_gen_prl} in the context of intercell communication of synthetic gene oscillators via a small autoinducer molecule. In general, the mean-field coupling is applied in a network of dynamical systems where each oscillator is having equal chance of uniform interaction from all the oscillators. On the other hand,  there are various types of stable steady states, which may not be possible to detect analytically from linear stability analysis due to their multistable behavior. AD state never produces in identical coupled systems using simple diffusive interaction but OD states may generate by proper choice of initial conditions and linear stability analysis fails to characterize such OD states due to multi stability.  For such limitations, it is not possible to get any information about the stability of OD state against any non-small random perturbation from the state. Again, there exists Lyapunov function based approach \cite{lasal,lyap} as a process in determining the stability of different steady states locally as well as globally but unfortunately there is no systematic way to construct Lyapunov functions for high dimensional systems and it depends on the exact form of the governing system. So in order to do the present work we avoid such limitation and concentrating on this intriguing BS approach. Thus it is significant to quantify all the multi stable steady states by BS measure. The value of BS lies in $[0,1]$ and quantifies what amount stable a state is in probabilistic sense against the basin volume. With the help of this measurement all coupling dependent steady states (OD, NHSS) as well as coupling independent state (i.e. AD state) can be quantified. The effect of coupling strength on the variation of different stable states is quantified in BS framework.  In BS measure, we integrate the whole network with a large population of initial states and give some probabilistic measure with respect to those initial points in the state space. We obtain analytical conditions of stabilization of various steady states that show excellent matching with our numerical simulations. Using rigorous bifurcation analysis we verify the results obtained analytically and appraise them by BS approach. For our investigation, we take coupled paradigmatic Duffing-Holmes and chaotic Lorenz oscillator to check the validation of our BS approach for global and random networks.

\section*{Results}
We start with a network of coupled oscillators with the isolated dynamics of each node of the network is given by $\dot {X}=F(X),$ where $X$ is a $m$-dimensional vector of the dynamical variables and $F(X)$ is the vector field. The general framework of coupled network is given by the following equation:
\begin{equation}
	\begin{array}{lcl}
		\dot{X_i}=F(X_i)+\epsilon \sum_{j=1}^{j=N}C_{ij}H(X_i, X_j), \;\;\;\;\;\; i=1,2,...,N,
	\end{array}
\end{equation}
where $N$ is the total number of nodes in the network, $\epsilon$ is the coupling strength, $C_{ij}$ are the elements of connectivity matrix and $H(X_i, X_j)$ is the coupling function between $i-$th and $j-$th node.

\subsection*{Duffing-Holmes oscillator}
We first consider a two-dimensional physical example, namely Duffing-Holmes (DH) oscillator \cite{dhott}:
\begin{equation}
	\begin{array}{lcl}
		\ddot{x}+b\dot{x}-x+x^3=0.
	\end{array}
\end{equation}
The oscillator has three steady states, namely two symmetrical stable steady states $(\pm 1,0)$ (which are spiral or node depending on the damping coefficient $b>0$) and a saddle point at $(0,0)$ irrespective of the values of the parameter $b$. For $b<0,$ each individual DH oscillator exhibit oscillatory state.  Recently, Tama\v{s}evi\v{c}i\={u}t\.{e} et al.  \cite{sd} discussed the stabilization of saddle fixed points of an uncoupled DH oscillator using modified unstable filter \cite{uf} method. The proposed technique is applicable only for $b>0$ where the DH oscillator is either stable node or spiral.  But they did not discuss the stabilization of saddle point in coupled oscillators. Here we study the stabilization of saddle point of coupled systems by taking all values of damping parameter $b$. In this context, detection and controlling both saddle and nonsaddle types of unstable steady states in high-dimensional nonlinear dynamical systems  based on fast-slow manifold separation and Markov chain theory is articulated in \cite{rev}.  
\par
\begin{figure}[ht]
		\centerline{
			\includegraphics[scale=0.48]{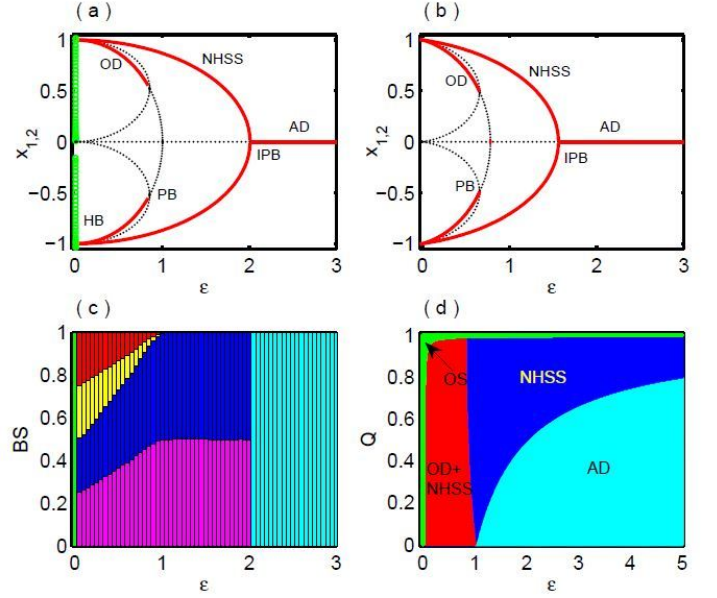}}
	\caption{  Two coupled Duffing-Holmes oscillators: bifurcation diagram with respect to coupling strength $\epsilon$ for (a) b= - 0.01, (b) b=0.5 where extreme values of $x_1$ and $x_2$ are plotted with coupling strength for $Q=0.5$. Red lines correspond for stable steady states, black dotted points are unstable steady states and green circle for oscillation state. PB: pitchfork bifurcation, OD: oscillation death, AD: amplitude death, IPB: inverse pitchfork bifurcation. (c)  Variation of BS for different values of coupling strength $\epsilon$. The color green stands for BS of oscillatory state, red and yellow for BS of stable IHSS states $E_{3,4}$, blue and magenta for BS of stable NHSS states $E_{1, 2}$ and color cyan correspond to BS of the HSS state $E_0$.  Other parameters: $b=-0.01, Q=0.5$. (d) Two parameter bifurcation diagram in the $\epsilon-Q$ plane where green, red, blue and cyan regions correspond to oscillatory state, coexistence of stable IHSS (OD) and NHSS, stable NHSS state and AD state respectively for $b=-0.01.$ }
	\label{two_coup_DH}
\end{figure}

 We consider globally coupled network through mean-field in the following form:
\begin{equation}
	\begin{array}{lcl}
		\dot x_i=y_i+\epsilon(Q\frac{\sum_{j=1}^{N}A_{ij}x_j+x_i}{d(i)+1}-x_i),\\
		\dot y_i=x_i-x_i^3-by_i+\epsilon(Q\frac{\sum_{j=1}^{N}A_{ij}y_j+y_i}{d(i)+1}-y_i),
	\end{array}
\end{equation}
for $i=1,...,N$. Here $\epsilon$ is the mean-field coupling strength, $d(i)$ is the degree of the  $i$-th node and $Q (0 \le Q < 1)$ is the mean-field density parameter. This mean-field density parameter $Q$ gives an additional free parameter that control the mean-field dynamics while $Q \rightarrow 0$ represents self-feedback case and $Q \rightarrow 1$ indicates the maximum mean-field density.  The elements of the connectivity matrix $A_{ij}=1$ if $i-$th and $j-$th  nodes are connected and zero otherwise. At first we consider a minimal network of two ($N=2$)  coupled Duffing-Holmes oscillators with mean field coupling and identify the parameter region for stabilized saddle point at origin. The coupled DH oscillator has a trivial steady state $E_0=(0, 0, 0, 0)$ which is the HSS solution of the system and the other four coupling dependent steady states: non-trivial homogeneous steady state (NHSS) $E_{1,2}=(\pm\alpha,\pm\beta,\pm\alpha,\pm\beta),$ and inhomogeneous steady state (IHSS) $E_{3,4}=(\pm\gamma,\pm\delta,\mp\gamma,\mp\delta)$ where $\beta=\epsilon(1-Q)\alpha$, $\alpha=\sqrt{1-b\epsilon(1-Q)-\epsilon^2(1-Q)^2}$, $\delta=\epsilon\gamma$, $\gamma=\sqrt{1-\epsilon(b+\epsilon)}.$ The characteristic equation corresponding to the fixed point $E_0$ is $[(\lambda+\epsilon)^2+b(\lambda+\epsilon)-1][(\lambda+\epsilon)^2+(b-2Q\epsilon)(\lambda+\epsilon)+\epsilon^2Q^2-2Qb\epsilon-1]=0.$ Using Routh Hurwitz(RH) criterion the saddle point $E_0$ is stable if $\epsilon>\frac{-b+\sqrt{b^2+4}}{2(1-Q)}$ and that stabilization of saddle point occurred through inverse pitchfork bifurcation. By performing the stability analysis we analytically obtain the inverse pitchfork bifurcation (IPB) point at the coupling strength $\epsilon_{IPB}=\frac{-b+\sqrt{b^2+4}}{2(1-Q)}$. From linear stability analysis we also analytically derive the Hopf bifurcation (HB) point at $\epsilon_{HB}=\frac{-b}{2(1-Q)}$ where up to this critical value of the coupling strength, coupled systems exhibit oscillatory states (Fig.~\ref{two_coup_DH}(a)). Further increment of the coupling strength leads to co-existence of IHSS and NHSS up to a certain threshold of interaction strength $\epsilon_{PB}=\frac{-b(2+Q)+\sqrt{b^2(2+Q)^2+8(-Q^2+2Q+2)}}{2(2+2Q-Q^2)}$ and after $\epsilon_{PB}$, IHSS are completely eliminated and only NHSS sustained up to $\epsilon_{IPB}$. So, using linear stability analysis and combining the  above results, structurally different dynamical states occur: AD exist for $\epsilon>\epsilon_{IPB}$, IHSS and NHSS (OD) coexist for $\epsilon_{HB}<\epsilon<\epsilon_{PB}$ and only NHSS exist for $\epsilon_{PB}<\epsilon<\epsilon_{IPB}$.

\par For numerical simulation, we choose the damping coefficient $b = -0.01$ for which an isolated oscillator exhibits oscillatory dynamics. At lower value of coupling strength $\epsilon$, four coupling dependent fixed points (i.e. NHSS and IHSS) that arise through Hopf bifurcation at $\epsilon=\epsilon_{HB}$, are stable. But as $\epsilon$ increases, two of these stable steady states $E_{3,4}$ become unstable at $\epsilon_{PB}$ and $E_{12}$ remains stable for the value of $\epsilon$ upto $\epsilon_{IPB}$ . At $\epsilon_{IPB}$, saddle point $(0,0,0,0)$ turns stable through IPB and remains stable for $\epsilon>\epsilon_{IPB}$. The corresponding bifurcation diagram (using XPPAUT \cite{xppaut}) is shown in Fig.~\ref{two_coup_DH}(a). Figure~\ref{two_coup_DH}(b) shows the bifurcation diagram with respect to coupling strength $\epsilon$ when $b=0.5$, for which an isolated oscillator approaches either to the steady state $(1, 0)$ or to $(-1, 0)$ where for negative values of $b$ different coupling dependent stable steady states appear through oscillatory states as shown in Fig.~\ref{two_coup_DH}(a). Here again, due to the introduction of coupling $\epsilon$, above mentioned four fixed points $E_{1,2}$ and $E_{3,4}$ become stable but $E_{3,4}$ remain stable only upto $\epsilon_{PB}$. Further increment in the value of $\epsilon$ makes the saddle point $(0,0,0,0)$ stable through an inverse pitchfork bifurcation at $\epsilon_{IPB}$. Figure~\ref{two_coup_DH}(c) shows how the BS of the steady states $E_{1,2}$ and $E_{3,4}$ change for different values of $\epsilon$. As can be seen, initially after the occurrence of Hopf bifurcation at $\epsilon=\epsilon_{HB}$, all the fixed points  ($E_{1,2,3,4}$) are equally probable although the probabilistic dominance of  $E_{3,4}$ are shrinking gradually whereas $E_{1,2}$ acquire more and more space in the basin volume. Such changes on   BS measure of $E_{3,4}$  gives a hint of the annihilation of $E_{3,4}$ which finally occurs at $\epsilon=\epsilon_{PB}$ where $E_{1}$ and $E_{2}$ share the basin with equal probability. But at $\epsilon=\epsilon_{IPB}$, BS of these steady states abruptly decrease to zero without any presage and further increase of $\epsilon$, the basin volume is fully covered by this HSS shown by cyan color with maximum BS i.e., $1$. From this figure, we  conclude that the BS for multi stable states (i.e. OD and NHSS) change with the variation of mean-field coupling strength $\epsilon$ while the BS for monostable state i.e. AD state remains unchanged with the variation of $\epsilon$. Therefore, the trend in the changes of the percentage	of the basin volume gives us a clear idea how the different steady states are evolving in a coupled system and which states will dominate the system and which	will disappear early. We also obtain similar results on stabilization of saddle point in two coupled DH oscillators when they are coupled through cross mean-field type configuration (see Supplementary Information section I). Figure~\ref{two_coup_DH}(d) represents the parameter region in $\epsilon-Q$ plane where green, red, blue and cyan regions respectively resembles the oscillatory state, coexistence of stable IHSS (OD) and NHSS, stable NHSS state and AD state for $b=-0.01$. For increasing values of $\epsilon$ firstly the coupling dependent fixed points get stabilized for almost all the values of $Q$ below the Hopf bifurcation curve $\epsilon=\frac{-b}{2(1-Q)}$. Then the saddle point $E_0$ becomes stable resulting in AD below the inverse pitchfork bifurcation curve $\epsilon=\frac{-b+\sqrt{b^2+4}}{2(1-Q)}$. 
\par We know that the presence of noise is common in real systems. To study the impact of noise in the steady states  we use additive Gaussian noise in the system and find that systems  still evolve around the steady states with small fluctuations which further implies that BS of each fluctuated steady states does not alter or vanish in the presence of noise. For detailed numerical observations see the Supplementary Information section II.

\begin{figure}[ht]
		\centerline{
			\includegraphics[scale=0.46]{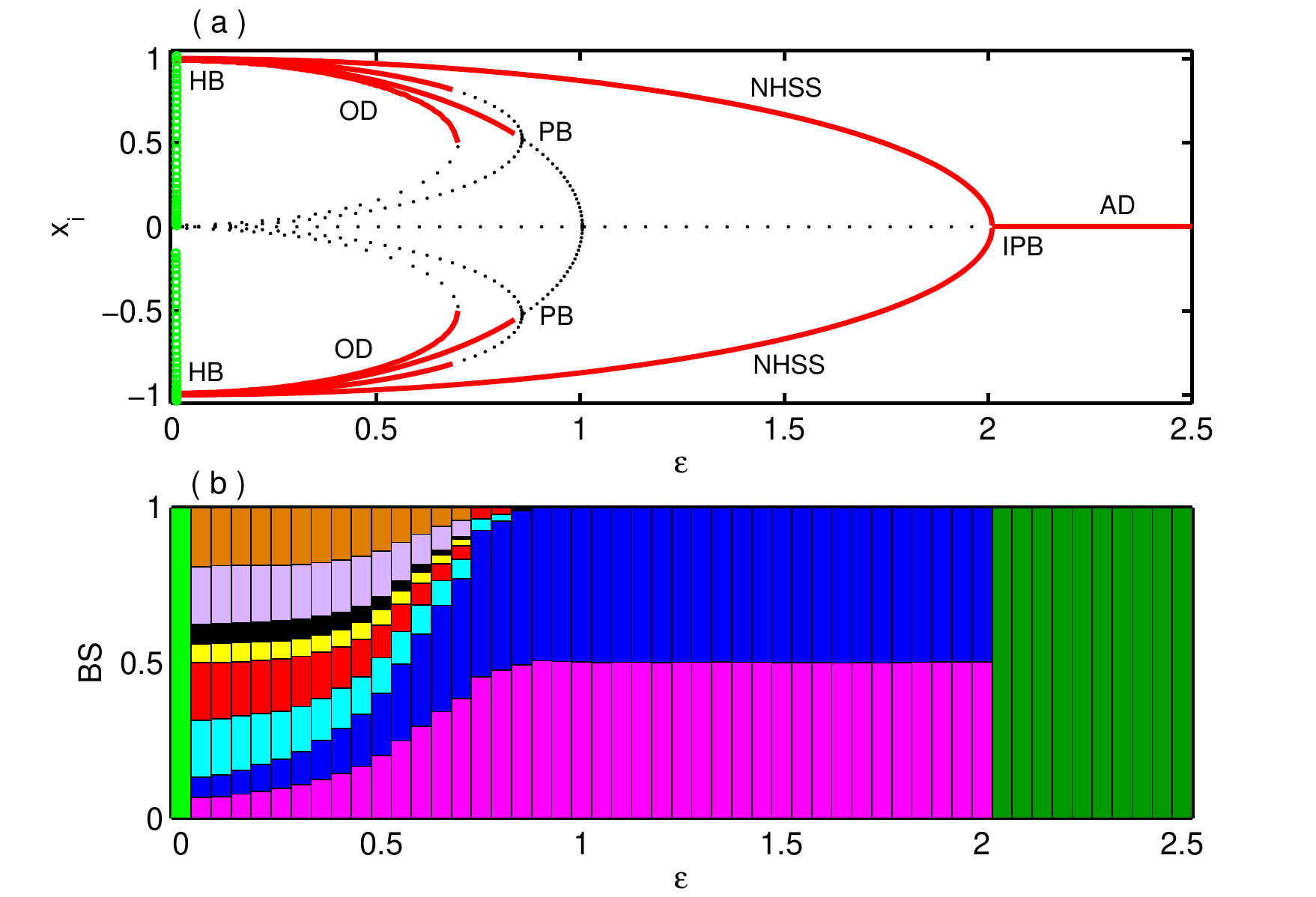}}
	\caption{ Four coupled Duffing-Holmes oscillators: (a) bifurcation diagram with respect to coupling strength $\epsilon$ for $b= - 0.01$ and $Q=0.5$, where extreme values of $x_i (i=1, 2, 3, 4)$  are plotted with coupling strength. Red lines correspond to stable steady states, black dotted points are unstable steady states and green circle for oscillation state.  (b) Variation of BS for different values of coupling strength $\epsilon$. The color green is for BS of oscillatory state, blue and magenta for BS of stable NHSS states, deep green for BS of stable saddle point $E_0$ (AD state) and other colors correspond to BS of different stable IHSS states. }
	\label{four_coup_DH}
\end{figure}

{\subsection*{Global network of Duffing-Holmes oscillators}}
Next we check the stabilization of saddle point in a network of equation (3) for higher values of $N>2$.  At first we start with a complete graph of size N. Fixed points of $N$ coupled oscillators are $E_0=(0,0,...,0)$, $E_{1,2}=(\pm\alpha,\pm\beta,\pm\alpha,\pm\beta,...,\pm\alpha,\pm\beta)$, and $E_{3,4}=(\pm\gamma,\pm\delta,\mp\gamma,\mp\delta,...,\pm\gamma,\pm\delta,\mp\gamma,\mp\delta)$ (for even number oscillators) where $\alpha, \beta, \gamma$ and $\delta$ are same as above. The fixed points $E_{1,2}$ are same for any choice of $N$ whereas $E_{3,4}$ are same only for even number of $N$.  Characteristic equation at $E_0$ is
\begin{equation}
	\begin{array}{lcl}
		[(\lambda+\epsilon)^2+b(\lambda+\epsilon)-1]^{N-1}[(\lambda+\epsilon)^2+(b-2Q\epsilon)(\lambda+\epsilon)+\\\epsilon^2Q^2-2Qb\epsilon-1]=0. 
	\end{array}
\end{equation}  
The distinct eigenvalues and critical bifurcation points of globally connected network (3) are same as for two coupled oscillators.
\par Next, we consider $N=4$ i.e. four globally coupled DH oscillators via mean-field coupling and the results are shown in Fig.~\ref{four_coup_DH}. Analytically it is not easy to  calculate all the coupling dependent fixed points (i.e. IHSS and NHSS), using bifurcation diagram (performed in XPPAUT \cite{xppaut}) we identify all the fixed points and by BS measurement we measure the amount of their stability for different values of coupling strength. In Fig.~\ref{four_coup_DH}(a), bifurcation diagram for the variables $x_i$ with respect to the coupling strength $\epsilon$ is plotted. For small values of $\epsilon$, through Hopf bifurcation at $\epsilon_{HB}=\frac{-b}{2(1-Q)}$, eight coupling dependent fixed points are stable. But as $\epsilon$ is increased, firstly six of these stable fixed points lose their stability through PB at $\epsilon_{PB}=\frac{-b(2+Q)+\sqrt{b^2(2+Q)^2+8(-Q^2+2Q+2)}}{2(2+2Q-Q^2)}$ and only two retain their stability. Even more increment in $\epsilon$ makes the two fixed points unstable and the saddle point (i.e., the origin) becomes stable through IPB. 
 The process of stabilization and destabilization of all the coupling dependent fixed points are clarified in terms of their BS which validates the whole mechanism in global scale. Figure ~\ref{four_coup_DH}(b) shows the variation of BS for different steady states by varying the mean-field coupling strength $\epsilon$. 
	As mentioned earlier, the blue and magenta color in Fig. ~\ref{four_coup_DH}(b) belong to class NHSS and they acquire more and more space in the basin if we increase the coupling strength continuously. On the other hand, the other cluster belonging to IHSS (six states in three symmetric groups) losing their stability and finally all of them vanish at $\epsilon=\epsilon_{PB}$ point. Further changes in $\epsilon$ makes those two fixed points (NHSS) equally probable in the basin i.e. each of them acquires half of the whole basin  and they become unstable at the point $\epsilon=\epsilon_{IPB}$. Then the saddle point (i.e. the origin) becomes stable for all points in the basin of attraction i.e. the basin volume is fully covered by this HSS.
 
 \begin{figure}[ht]
 		\centerline{
 			\includegraphics[scale=0.46]{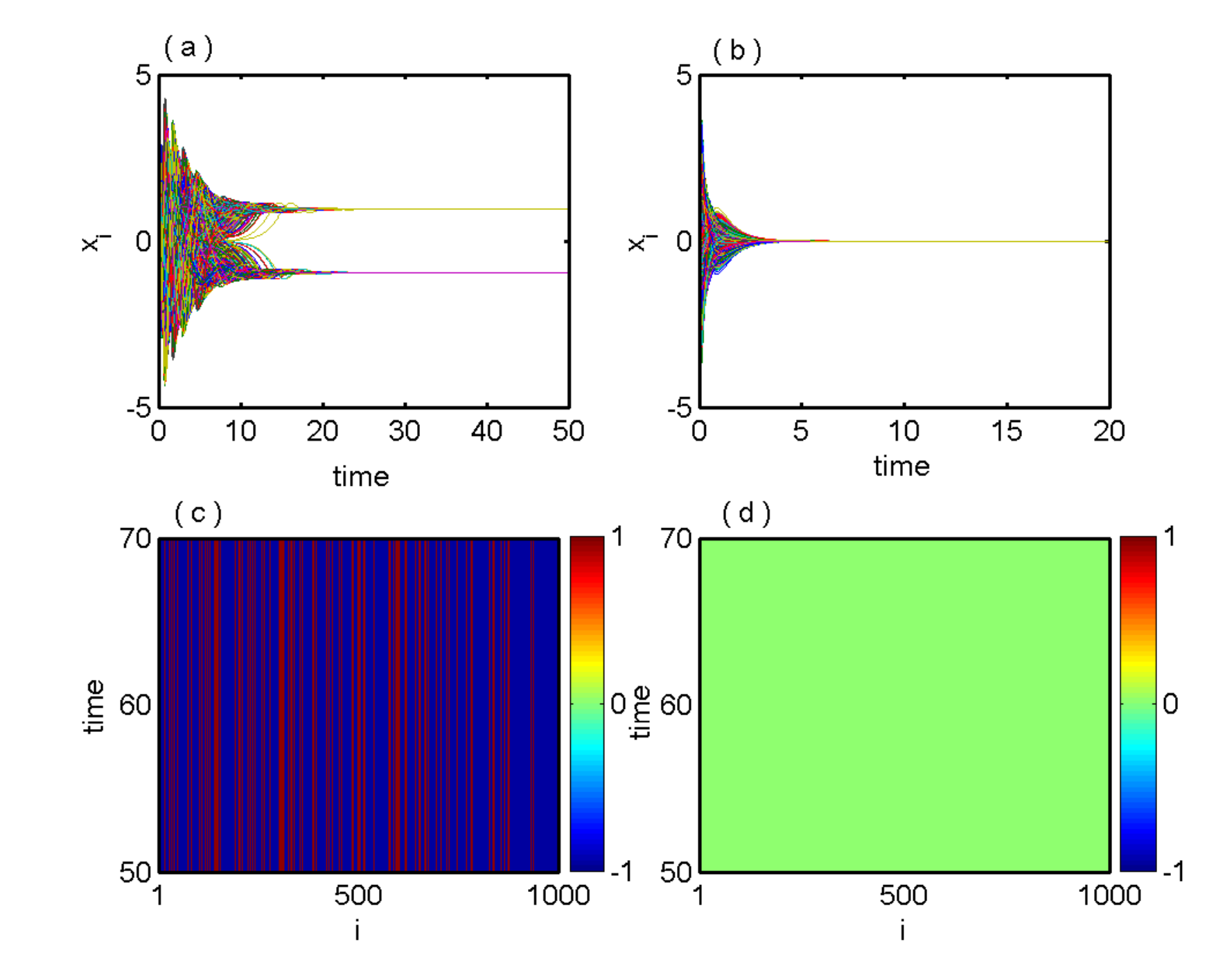}}
 	\caption{ Globally coupled Duffing-Holmes oscillators for $N=1000$: (a) time series of $x_i, i=1, 2, ..., 1000$ show the IHSS state for $\epsilon=0.3$. (b) Time series of stabilized saddle point $E_0$ for $\epsilon=2.2$. (c) and (d) corresponding space-time plot of (a) and (b) respectively showing stable IHSS and HSS states. Other parameters are: $Q=0.5, b= - 0.01$. }
 	\label{1000_coup_DH}
 \end{figure} 

\par Next we will verify numerically whether the stabilization of saddle and all coupling dependent steady states using the proposed coupling scheme is working in a large network. For our case, we choose $N=1000$ globally coupled DH oscillators via mean-field, the analyzed results are illustrated in Fig.~\ref{1000_coup_DH}. Figure~\ref{1000_coup_DH}(a) shows time series of  $x-$components of all the $1000$ oscillators with $\epsilon=0.3$ that depicts the stabilization of the IHSS resulting in OD. For larger value of $\epsilon$ ($\epsilon=2.2$), all the oscillators populate to a single steady state, that is, the saddle point (the origin) gets stabilized, time series are shown in Fig.~\ref{1000_coup_DH}(b). The corresponding space-time plots are shown in Fig.~\ref{1000_coup_DH}(c) and Fig.~\ref{1000_coup_DH}(d) respectively. The parameter space in $\epsilon-Q$ plane for global network is same as in Fig. 1(d) for two coupled DH oscillators, as the distinct eigenvalues of the characteristic equation (4) are identical with two coupled case but with different multiplicity.  \\

\subsection*{Random network of Duffing-Holmes oscillators}
\begin{figure}[ht]
		\centerline{\includegraphics[scale=0.42]{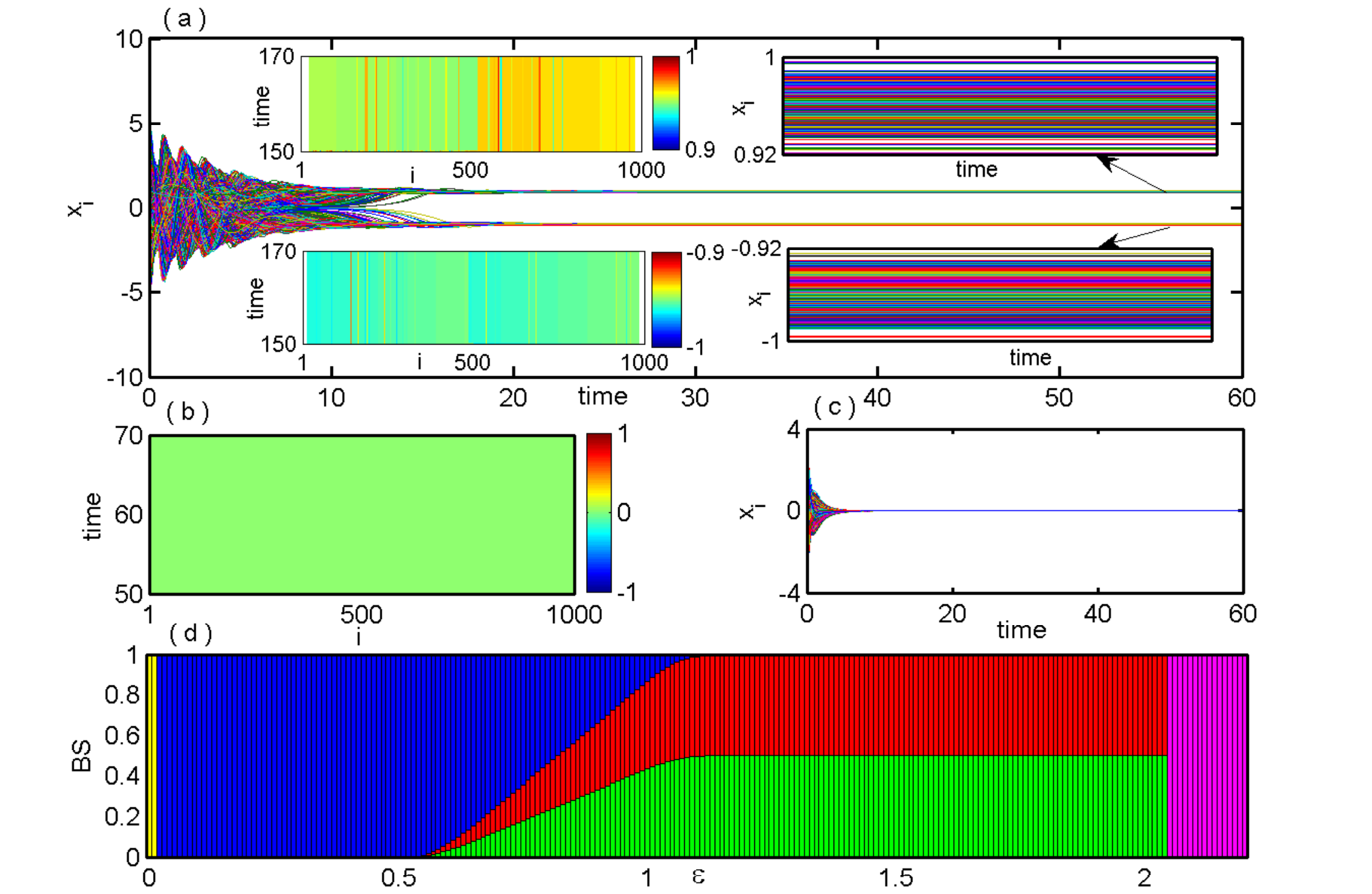}}
	\caption{ Randomly coupled Duffing-Holmes oscillators ($N=1000$): (a) time series of $x_i, i=1, 2, ..., 1000$ show the MCOD state for $\epsilon=0.3$. Right and left inset figures in (a) show the time series of coupling dependent different steady stables and corresponding spatio-temporal plots respectively. (b) Space-time plot and (c) corresponding time series of stabilized saddle point $E_0$ for $\epsilon=3.0$. (d) BS of MCOD, NHSS and AD states against the coupling strength $\epsilon$. The oscillatory state, MCOD, NHSS and AD states are represented by yellow, blue, red / green and magenta colors respectively. Other parameters are: $Q=0.5, b= - 0.01$.}
	\label{random_coup_DH}
\end{figure}

In this section we are concerned with the phenomenon of stabilization of saddle point in Erd\H{o}s-R\'{e}nyi random networks of DH oscillators and the results are shown in Fig.~\ref{random_coup_DH} where the probability of existence of an edge between any two vertices of the random network is taken as $p=0.01$. Figure~\ref{random_coup_DH}(a) shows the time series of the $x-$components of all the $N=1000$ oscillators characterizing MCOD state for $\epsilon=0.3$. The inset figures (right panel) show the magnified time-series plots for better visibility of the MCOD state. The space-time plots corresponding to these time-series are given in the insets (left panel) of Fig.~\ref{random_coup_DH}(a). Figures~\ref{random_coup_DH}(b) and \ref{random_coup_DH}(c) show the space-time plot  and the corresponding time series represent stabilization of the saddle point (the origin) resembling AD state for $\epsilon=3.0$. Figure~\ref{random_coup_DH}(d) depicts the variation of the BS of the MCOD and NHSS states for the random network. Due to failure of calculation of all the MCOD states analytically in a random network, we consider all the states as MCOD state and represented by blue color in Fig.~\ref{random_coup_DH}(d).  After the Hopf bifurcation, MCOD state dominates over the NHSS state where relative acceptance of MCOD in BS measure is almost unity and the probability of occurrence of NHSS is almost nil. With increasing the values of the coupling strength $\epsilon$, the probability of getting NHSS states increases and MCOD state decreases.  Then the BS of NHSS starts increasing gradually and vanishing of BS of the MCOD state is observed for $\epsilon\simeq 1.24$. NHSS remains stable further upto $\epsilon\simeq 2.05$ from where the saddle point becomes stable through IPB with BS unity.

\begin{figure}
	{\includegraphics[height=6.5cm,width=9.0cm]{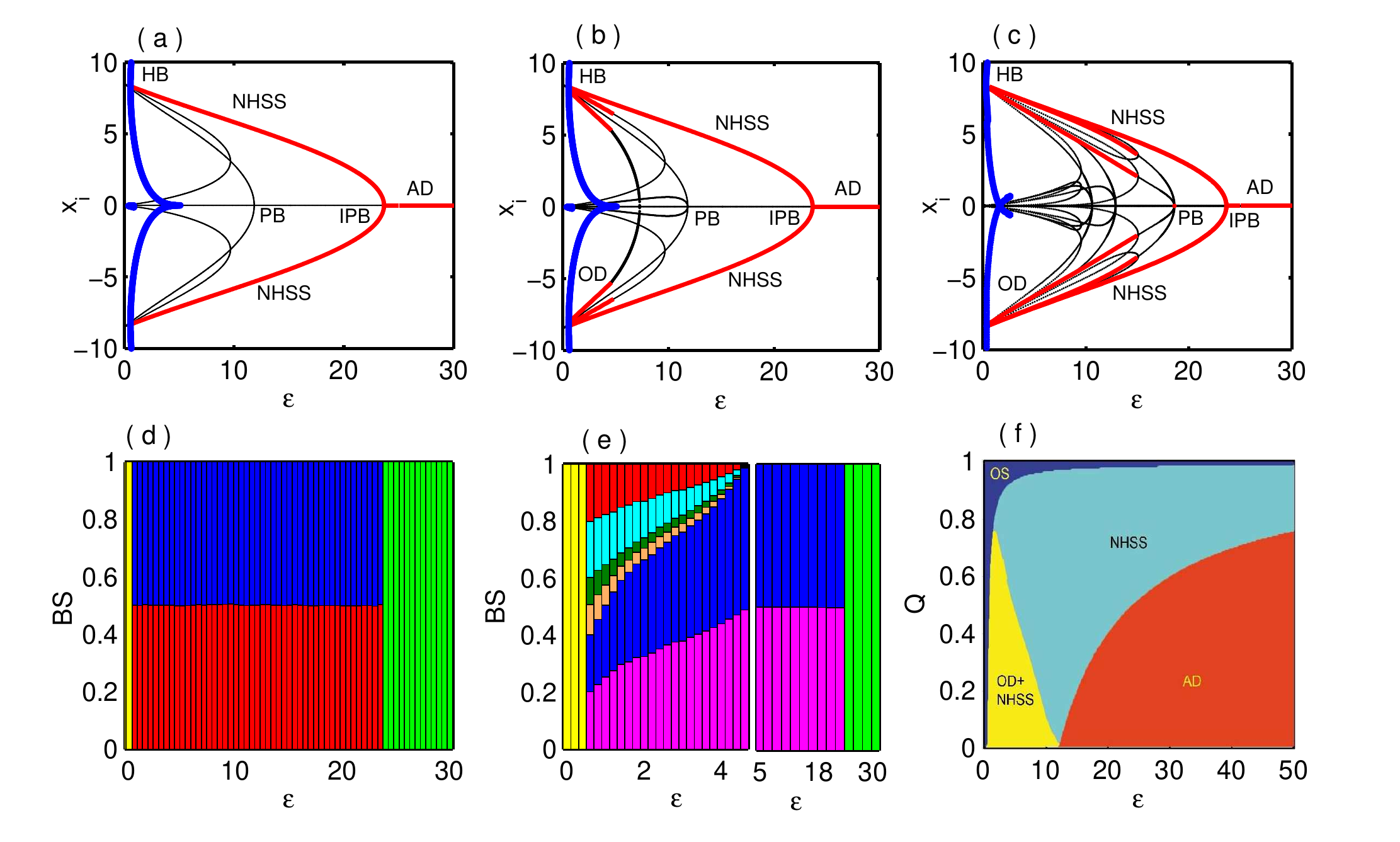}}
	\caption{Coupled Lorenz oscillators: Bifurcation diagrams by changing the coupling strength $\epsilon$ for (a) $N=2$ and (b) $N=4$ globally coupled oscillators. (c) Bifurcation diagram for randomly coupled $N=4$ oscillators, (d) For $N=2$ and (e) $N=4$ globally coupled oscillators, the variation of BS with respect to coupling strength $\epsilon.$ Other parameter $Q=0.5$. (f) Parameter region in $\epsilon-Q$ plane for $N=4$ globally coupled network.}
	\label{lorenz24}
\end{figure}

{\subsection*{Lorenz oscillators}}
For quantifying the different stable steady states using BS measure, we extend our investigation on coupled paradigmatic chaotic Lorenz oscillator \cite{lorenz}. We consider $N$ Lorenz oscillators interacting through mean-field diffusive coupling. The mathematical equations of the coupled systems are described as: 
\begin{equation}
	\begin{array}{lcl}
		\dot x_i=\sigma(y_i-x_i)+\epsilon(Q\frac{\sum_{j=1}^{N}A_{ij}x_j+x_i}{d(i)+1}-x_i),\\
		\dot y_i=rx_i-y_i-x_iz_i+\epsilon(Q\frac{\sum_{j=1}^{N}A_{ij}y_j+y_i}{d(i)+1}-y_i),\\
		\dot z_i=x_iy_i-bz_i,
	\end{array}
\end{equation}
for $i=1,2,...,N$. In absence of coupling term, each oscillators oscillate chaotically for $\sigma=10,r=28$ and $b=\frac{8}{3}$ and the individual systems have a saddle fixed point at origin and two unstable fixed point at $(\pm\sqrt{b(r-1)},\pm\sqrt{b(r-1)},r-1)$. Here $\epsilon$ and $Q$ are the coupling strength and mean-field density parameter respectively.\\
For $N=2$, the fixed points are $E_0=(0,0,0,0,0,0),E_{1,2}=(\pm\alpha^*,\pm\beta^*,\gamma^*,\pm\alpha^*,\pm\beta^*,\gamma^*)$, where $\alpha^*=\sqrt{b\frac{r\sigma-(1+(1-Q)\epsilon)(\sigma+(1-Q)\epsilon)}{\sigma+(1-Q)\epsilon}}$, $\beta^*=\frac{\alpha^*}{\sigma}(\sigma+(1-Q)\epsilon)$ and $\gamma^*=\frac{\alpha^*\beta^*}{b}$.   The characteristic equation at $E_0$ is 
\begin{equation}
	\begin{array}{lcl}
		(\lambda+b)^2[\lambda^2+\{\epsilon(1-Q)+(1+\sigma)\}\lambda+\\(1+\epsilon-Q\epsilon)(\sigma+\epsilon-Q\epsilon)-r\sigma]\times\\
		~~ [\lambda^2+(2\epsilon+1+\sigma)\lambda+(1+\epsilon)(\sigma+\epsilon)-r\sigma]=0
	\end{array}
\end{equation}
The trivial fixed point $E_0$ is stable through inverse pitchfork bifurcation at $\epsilon_{IPB}=\frac{-(1+\sigma)+\sqrt{4r\sigma+(\sigma-1)^2}}{2(1-Q)}.$     
\\For fixed values of the above system parameters, from eigenvalue analysis the NHSS points $E_{1,2}$ becomes stable for $\frac{0.2791}{1-Q}<\epsilon<\frac{-(1+\sigma)+\sqrt{4r\sigma+(\sigma-1)^2}}{2(1-Q)}$. 
The results are shown in  Fig.~\ref{lorenz24}. Figures~\ref{lorenz24}(a) and (b) show bifurcation diagrams with respect to $\epsilon$ for $N=2$ and $N=4$ respectively with $Q=0.5$ fixed. As in Fig.~\ref{lorenz24}(a), due to the  presence of coupling, two stable fixed points $E_{1,2}$ develop together with six unstable fixed points through Hopf bifurcation at $\epsilon_{HB}=\frac{0.2791}{1-Q}$ and $E_{1,2}$ remain stable for $\epsilon$ upto $\epsilon_{IPB}=\frac{-(1+\sigma)+\sqrt{4r\sigma+(\sigma-1)^2}}{2(1-Q)}$. The saddle point $E_0$ becomes stable through an inverse pitchfork bifurcation at $\epsilon_{IPB}$, and persists for any higher values of $\epsilon$ as well.  For $N=4$, Fig.~\ref{lorenz24}(b) shows that immediately after the occurrence of Hopf bifurcation at $\epsilon_{HB}=\frac{0.2791}{1-Q}$, six coupling dependent stable fixed points (comprising of both IHSS and NHSS states) emerge together with six unstable fixed points. But among them, the fixed points except the NHSS $E_{1,2}$ lose their stability soon and only $E_{1,2}$ remain stable for higher values of $\epsilon$. Similarly as before, through IPB at $\epsilon=\epsilon_{IPB}$, $E_1$ and $E_2$ collides and $E_0$ turns stable. Figure~\ref{lorenz24}(c) shows the bifurcation diagram against $\epsilon$ for a network of $N=4$ randomly connected nodes where the appearance of six coupling dependent stable fixed points along with many other unstable fixed points can be seen. Similarly as in the previous cases, here also $E_{1,2}$ retain their stability for higher values of $\epsilon$ than the others and lose it stability at $\epsilon_{IPB}$ and further higher coupling strength promotes the entire systems to the AD state. Figures~\ref{lorenz24}(d) and ~\ref{lorenz24}(e) measure all the stable steady states that appear for $N=2$ and $N=4$ respectively in terms of their BS. Figure~\ref{lorenz24}(d) shows that the BS of both $E_1$ and $E_2$ are non-zero and more or less the same for all values of $\epsilon$ upto $\epsilon_{IPB}$. As $\epsilon$ increases further, BS of $E_1$ and $E_2$ turns into zero and BS of $E_0$ becomes unity. On the other hand, soon after the Hopf bifurcation all the six coupling dependent stable fixed points get non-zero BS but $E_1$ and $E_2$ have larger BS than the others, as in Fig.~\ref{lorenz24}(e) (left part). Increasing $\epsilon$, BS of the other fixed points become zero and $E_1$ and $E_2$ shares almost the same BS value upto $\epsilon=\epsilon_{IPB}$. After that BS of both $E_{1,2}$ becomes zero and that of $E_0$ appears to be $1$ (right part in Fig.~\ref{lorenz24}(e)). 

\par Finally, Fig.~\ref{lorenz24}(f) depicts the  parameter region in $\epsilon-Q$ plane for globally coupled $N=4$ oscillators. Here blue, yellow, cyan and red regions signify oscillatory state, co-existence of OD and NHSS states, stable NHSS state and AD state (i.e., the stabilization of saddle $E_0$) respectively. The oscillatory state (blue region) and coexistence of OD and NHSS (yellow region) or stable NHSS (cyan region) are separated by the Hopf bifurcation curve $\epsilon=\frac{0.2791}{1-Q}$. From this curve it is clear that the oscillatory state persists for higher values of coupling strength $\epsilon.$ The stability of OD or NHSS loses when the value of $\epsilon$ passes through the inverse pitchfork bifurcation curve $\epsilon=\frac{-(1+\sigma)+\sqrt{4r\sigma+(\sigma-1)^2}}{2(1-Q)}$.

 \begin{figure}[ht]
 		\centerline{
 			\includegraphics[scale=0.46]{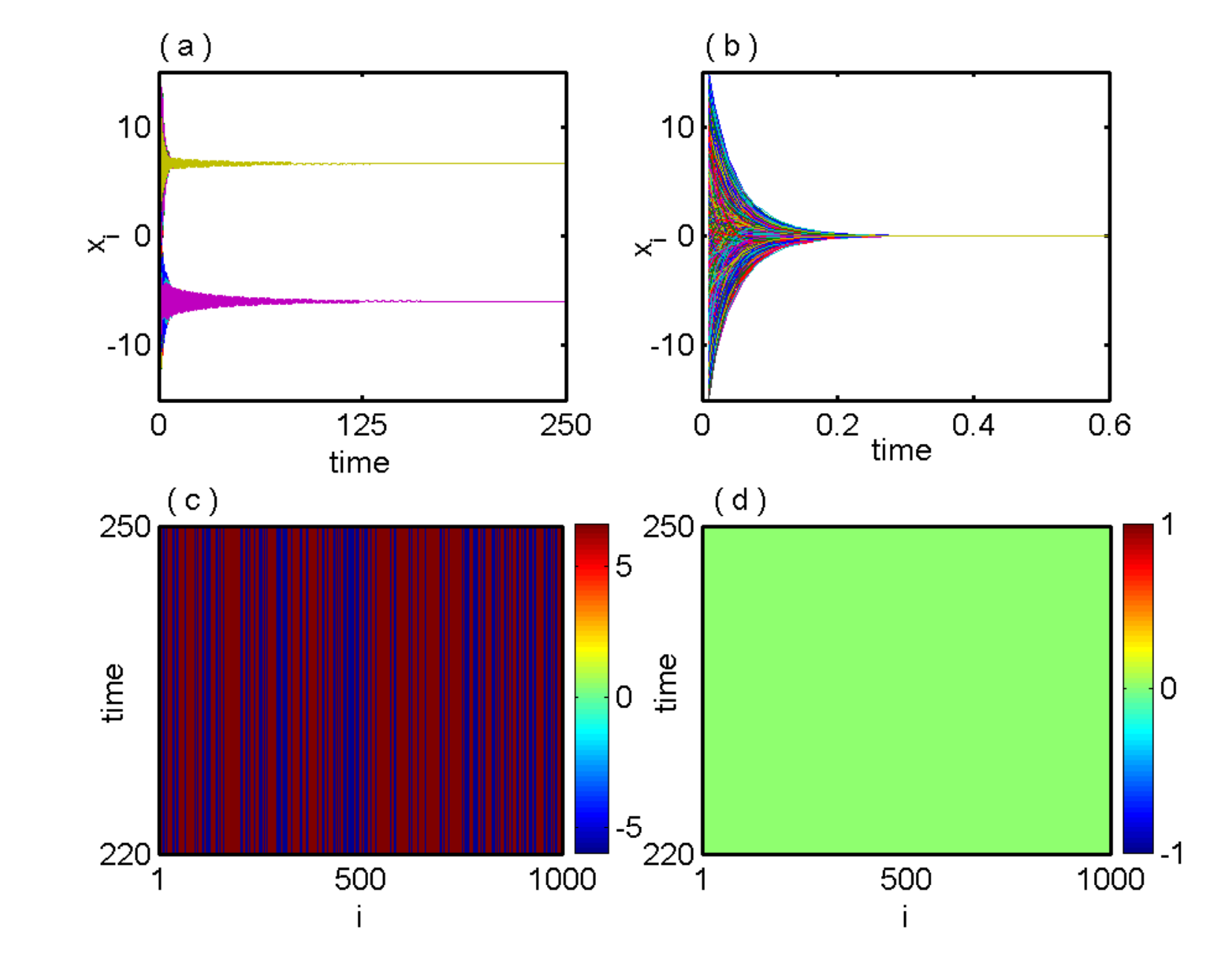}}
 	\caption{ Global network of Lorenz oscillators: (a) and (b) show the time series of IHSS and HSS states for coupling strength $\epsilon=5.5$ and $\epsilon=30$ and (c), (d) represents the corresponding space-time plot of (a) and (b) respectively. Here $N=1000$ and $Q=0.5$.}
 	\label{lorenz1000_global}
 \end{figure}

{\subsection*{Networks of Lorenz oscillators}}
Next we will explore the proposed coupling scheme is applicable for large number of chaotic oscillators. To quantify the stability of different steady states using BS measure in global and random network. The characteristic equation at $E_0$ of network (5) is
\begin{equation}
	\begin{array}{lcl} (\lambda+b)^N[\lambda^2+\{2\epsilon(1-Q)+(1+\sigma)\}\lambda+\\(1+\epsilon-Q\epsilon)(\sigma+\epsilon-Q\epsilon)-r\sigma]~\times\\
		~~~~~~[\lambda^2+(2\epsilon+1+\sigma)\lambda+(1+\epsilon)(\sigma+\epsilon)-r\sigma]^{N-1}=0.
	\end{array}
\end{equation} 

\par Taking a network of $N=1000$ globally coupled Lorenz oscillators with $Q=0.5$, the numerical results are shown in Fig.~\ref{lorenz1000_global}. Figure~\ref{lorenz1000_global}(a) shows time evolution of the $x-$components of all the $1000$ oscillators with $\epsilon=5.5$ that represents the stabilization of IHSS resembling OD. Whereas for $\epsilon=30$, the saddle point (origin) appears to be stable, time series shown in Fig.~\ref{lorenz1000_global}(b). Figures~\ref{lorenz1000_global}(c) and \ref{lorenz1000_global}(d) depict the corresponding space-time plots respectively.

\begin{figure}[ht]
		\centerline{
			\includegraphics[scale=0.46]{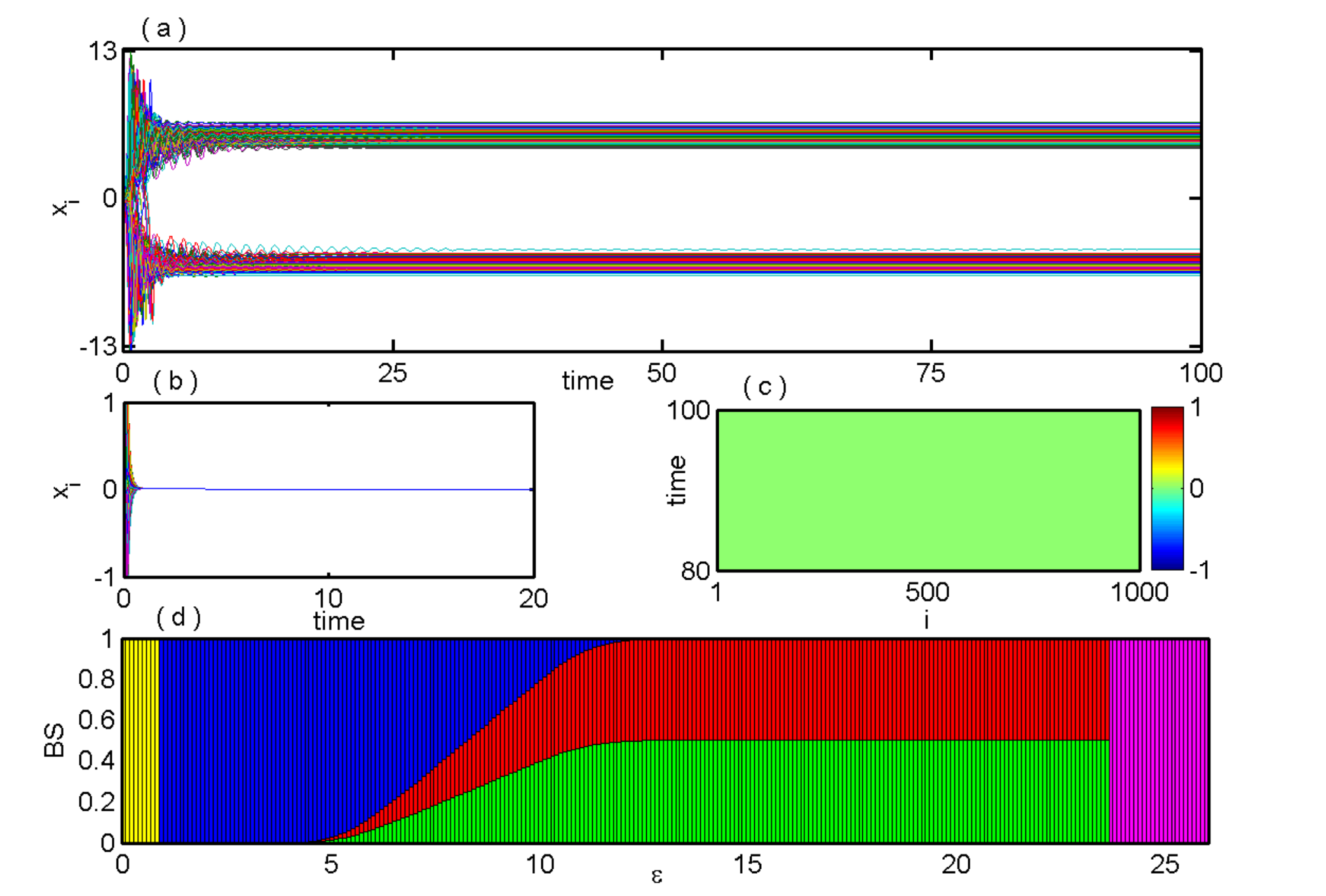}}
	\caption{ Random network of Lorenz oscillators: (a) Time series of $x_i, (i=1, 2, ..., 1000)$ shows MCOD state for $\epsilon=6.0.$ (b) Time series of $x_i, (i=1, 2, ..., 1000)$ show the stabilized saddle state for $\epsilon=30.0$ and (c) corresponding space-time plot. (d) Variation of BS with respect to the coupling strength $\epsilon$ where yellow color represents oscillatory behaviors, blue for MCOD, red and green for corresponding NHSS states and magenta for AD state.  Other parameter fixed at $Q=0.5$ and $N=1000$.}
	\label{lorenz1000_random}
\end{figure}
 
\par Results regarding MCOD state and saddle stabilization in  Erd\H{o}s-R\'{e}nyi random networks of coupled Lorenz systems are given in Fig.~\ref{lorenz1000_random}. Time series of the $x-$components of all the $N=1000$ oscillators revealing MCOD state for $\epsilon=6.0$ and $Q=0.5$ are shown in Fig.~\ref{lorenz1000_random}(a).  In Fig.~\ref{lorenz1000_random}(b) stable AD state ensuing after a concise transient window and Fig.~\ref{lorenz1000_random}(c) shows the corresponding space-time plots representing stabilization of the saddle point (the origin) reflecting AD for $\epsilon=30.0$ and $Q=0.5$.
The dependence of the BS of the MCOD and NHSS states on coupling strength $\epsilon$ for the random network of Lorenz systems is characterized in Fig.~\ref{lorenz1000_random}(d). Here, after the Hopf bifurcation at $\epsilon\simeq 0.9$, the MCOD state retains BS almost $1$ and the BS of NHSS is very small upto $\epsilon\simeq 4.3$. In fact, for $4.3<\epsilon<13.5$ the states MCOD and NHSS co-exist but then BS of NHSS develops with tantamount and that of the MCOD state vanishes at $\epsilon\simeq 13.5$ and NHSS remains stable further upto $\epsilon\simeq 23.7$ from where the saddle point becomes stable abruptly without any pre-warning and carries BS unity further.

\section*{Discussion}
\noindent In this work we have studied basin stability (BS) measure to quantify the stability of different stable steady states of coupled dynamical systems interacting through  mean-field coupling. BS is an universal concept to quantify the stability of governing dynamical systems under a non-uniform distribution of perturbations. Using mean-field coupling configuration, we have obtained a homogeneous stable steady state (i.e. AD state) which is inherently saddle equilibrium point of the individual oscillator and also showed  that the transition from inhomogeneous steady states (resembling OD) to homogeneous steady state (i.e. AD state) via stabilization of NHSS state. We identify the underlying mechanism to stabilize the saddle fixed points in a network of coupled oscillatory systems. The transition routes between different states of coupled systems are discussed through rigorous bifurcation analysis and confirmed with the obtained analytical results. We also map the different steady states in the wide parameter space by varying the mean-field coupling strength $\epsilon $ and mean-field density parameter $Q$. All the steady states are quantified by the value of BS. In contrast to this we found that the BS of OD states gradually decreases as coupling strength increases. After annihilation of the BS of multi-stable OD states, the BS of NHSS states become prevalent with almost equal ratio. But further increasing of coupling strength NHSS states become unstable without any presage and immediately AD state is stabilized. In the context of oscillations suppression studies, all the previous works have been done by considering the specific initial conditions in the phase space and no one examined the whole basin volume therefore ignoring the multistability nature of the steady states.  As multistable character is ubiquitous in natural systems so we clearly elucidate a global stability measure by means of basin stability. To validate the BS measure, we have considered a large number of initial states following \cite{BS-nat_phys}. All these phenomena and measures are performed using  smaller size of networks  (for N=2 and N=4) as well as  network of  bigger size (N=  1000).   We test our proposition and statistical measure not only in complete graph but also in random network. For both cases, our analytical and numerical simulations give  proper insight to track the multistablity features present in the systems.   The models considered here cover the characteristics of limit cycle (Duffing-Holmes oscillator) or chaotic attractor (Lorenz system) having hyperbolic fixed points.   There are many real systems such as laser \cite{laser2} and geomagnetic \cite{mpcps} which are modeled like Lorenz systems or mimic of Lorenz systems after some transformations and the results of our approach can be easily implemented. 
\par Our considered mean-field coupling is one of the most natural coupling scheme which is previously extensively applied to different branches of science and engineering.   This strongly means that our approach is not limited to a particular situation or for some particular systems, rather this mechanism is applicable in wide range of	systems throughout all these disciplines. 	 Also multistable feature is omnipresent in nature and widespread phenomenon in dynamical systems that appears in diverse fields ranging from physics, chemistry, biology to social systems \cite{ulrike}. There are numerous systems in which multistability	originates that include the human brain, semiconductor materials, chemical reactions, metabolic system, arrays of coupled lasers, hydrodynamical systems, various ecological systems, artificial and living neural systems etc. We believe that this study will broaden our understanding of stabilization of saddle points  in multistable dynamical networks where units are connected via mean-field. Further we have shown that the critical mean-field coupling strength is independent of the size of the network but only depends on the largest real part of the eigenvalue of individual oscillator (refers to Linear Stability Theorem in Method Section). 

\section*{Methods}
\noindent {\bf Basin Stability Measure:}\\
Let $I$ be the set of initial values for a given coupled system of $N$ oscillators which is a bounded subset of $R^N$. Suppose $X_k\in I$ is an asymptotically stable equilibrium point of the given system. Now let $B\subset I$ be the basin of attraction of the stable state $X_k$ (i.e, the solution of the system starting from any $z\in B$ asymptotically converges to $X_k$ as $t\rightarrow\infty$).  
\par We numerically integrate the given system for $V$ points which are drawn uniformly at random (sufficiently large) from $I$. Let $V_k$ be the count of the initial conditions that finally arrives at the stable steady state $X_k$. Then the BS for the fixed point $X_k$ is estimated as $\frac{V_k}{V}$.\\
\par \noindent {\bf Numerical Simulation:}\\
For numerical integration, we used fifth-order Runge-Kutta-Fehlberg algorithm with fixed step size $\Delta t=0.01$. For simulations of BS measure we choose sufficiently large number (for regular networks $20000$ and for irregular networks, $5000$) of initial conditions and all random initial conditions are chosen from  $ [-5,5]\times [-5,5] $ for coupled Duffing-Holmes oscillators and $[-20,20]\times [-30,30]\times [0,50]$ for coupled Lorenz oscillators.\\
\par \noindent {\bf Linear Stability Theorem:}\\
{\it {If $\dot {X}=f(X)$ be $m-$dimensional dynamical system which exhibits a saddle equilibrium point $O$, the saddle equilibrium point can be stabilized in globally mean-field coupled of N identical systems and the critical coupling strength is $k> k^*=\frac{\lambda^*}{1-Q}$, where $\lambda^*$ is the maximum real part of eigenvalues of the isolated system at the equilibrium point $O$ and $Q (0\le Q<1)$ is the mean-field density parameter.}}
\\
{\bf Proof:} Consider N identical systems interacting through global mean-field diffusive coupling as follows:
	\begin{center}
		$\dot{X_i}=f(X_i)+k(Q \bar{X} - X_i),\;\;\;\;\;\;   i=1,2,...,N,$
	\end{center}
	where $f(X_i)$ be the evolution equation of the $i^{th}$ system, $X_i$ denotes $m-$dimensional state vector, $k$ be the mean-field coupling strength, $Q$ is the mean-field density parameter and $\bar{X} = \frac{1}{N}\sum_{i=1}^{N} X_i.$
	\par The isolate system $\dot {X}=f(X)$ possess a saddle equilibrium point $O.$ So the Jacobian matrix $A=J_{X=O}$ of this system has at least two real eigenvalues with opposite sign.
	Let $\lambda^*$ be the maximum real part of eigenvalues $\lambda_1, \lambda_2,\lambda_3,...,\lambda_m.$
	\par The Jacobian matrix of the above coupled systems at the trivial equilibrium point $\underbrace{(O,O,...,O)}_{\normalfont{ N times}}$ is 
	\[ \left[ \begin{array}{cccc}
	A+(\frac{kQ}{N}-k)I_m & \frac{kQ}{N}I_m & .\;.\;. & \frac{kQ}{N}I_m \\
	\frac{kQ}{N}I_m & A+(\frac{kQ}{N}-k)I_m & \;.\;.\;. & \frac{kQ}{N}I_m \\
	.\:.\:. & .\:.\:. & .\:.\:. & .\:.\:. \\
	\frac{kQ}{N}I_m & \frac{kQ}{N}I_m & .\;.\;. & A+(\frac{kQ}{N}-k)I_m \end{array} \right]\]
	\par The corresponding \emph{characteristic equation} is
	$$det[A-k(1-Q)I_m-\lambda I_m].\{ det[A-kI_m-\lambda I_m] \}^{N-1} = 0.$$
	The eigenvalues are\\
	 ${\{\lambda^*-k(1-Q),\\\lambda_2-k(1-Q),\lambda_3-k(1-Q)},\cdot\cdot\cdot,$\\${\lambda_m-k(1-Q)\}}$ \\and ${\{\lambda^*-k,\lambda_2-k,\lambda_3-k,\;.\;.\;.\;,\lambda_m-k\}}$ $(N-1)$ times. The saddle point $O$ is stable if all the real parts of the eigenvalues are negative negative. For this it is sufficient to make $\lambda^*-k(1-Q)<0$. From this we have the critical coupling strength is $k^*=\frac{\lambda^*}{1-Q}$. \\\\


 \noindent \\ \textbf{Author contributions} \\
 S.R., B.K.B., S.M., C.H. and D.G. designed and performed the research as well as wrote the paper.\\
 
 {\large \bf Supplementary information:}
 
 \section{Cross mean-field Interaction}
 
 \noindent We briefly  discuss the different steady states in two coupled Duffing-Holmes oscillators interacting through cross mean-field coupling. We calculate the equilibrium points and derive the critical coupling strength using linear stability analysis. Numerical simulations confirm the analytical results and the probability of initial conditions for approaching different steady states using basin stability (BS) measure are calculated.  

 \par We consider two coupled Duffing-Holmes oscillators interacting through the cross mean-field coupling and the mathematical form  as follows: 
 \begin{equation}
 \begin{array}{lcl}
 \dot x_1=y_1+\epsilon(Q\frac{y_1+y_2}{2}-x_1),\\
 \dot y_1=x_1-x_1^3-by_1+\epsilon(Q\frac{x_1+x_2}{2}-y_1),\\
 \dot x_2=y_2+\epsilon(Q\frac{y_1+y_2}{2}-x_2),\\
 \dot y_2=x_2-x_2^3-by_2+\epsilon(Q\frac{x_1+x_2}{2}-y_2),
 \end{array}
 \end{equation}
 where $\epsilon$ is the cross mean-field coupling strength and $Q (0\le Q < 1)$ is the mean-field density parameter, as stated in the main text. The above coupled equation has the following fixed points:\\
 (i) trivial steady state $E_{0}=(0, 0, 0, 0)$ which is the homogeneous steady state (HSS) solution of the system,\\
 (ii) two coupling dependent steady states $E_{1, 2}=(\alpha, \beta, \alpha, \beta)$ where $\alpha=\frac{1+\epsilon Q}{\epsilon} \beta$ and $\beta=\pm \sqrt{\frac{\epsilon^2}{1+\epsilon Q}-\frac{(b+\epsilon)\epsilon^3}{(1+\epsilon Q)^3}}$. This steady state corresponds to the non-trivial homogeneous steady state (NHSS). The other coupling dependent steady state is\\
 (iii) $E_{3, 4}=(\gamma, \delta, -\gamma, -\delta)$ where $\gamma=\pm \sqrt{1-(b+\epsilon)\epsilon}$ and $\delta=\pm \epsilon \sqrt{1-(b+\epsilon)\epsilon}$. This state corresponds to the inhomogeneous steady state (IHSS).
 \par The eigenvalues corresponding to the steady state $E_{0}$ of the coupled system are
 $$\lambda_{1, 2}=\frac{-(b+2 \epsilon)\pm \sqrt{b^2+4}}{2}$$
 and 
 $$\lambda_{3, 4}=\frac{-(b+2 \epsilon)\pm \sqrt{b^2+4 \epsilon^2 Q^2+8\epsilon Q+4}}{2}$$
 From the eigenvalue analysis we derive the hopf bifurcation point (HB) at the coupling strength $\epsilon_{HB}=-\frac{b}{2}$, the inverse pitchfork bifurcation point (IPB) at the coupling strength $\epsilon_{IPB}=\frac{-(b-2Q)+\sqrt{(b-2Q)^2+4(1-Q^2)}}{2(1-Q^2)}$. The origin is stable if $\epsilon>\frac{-(b-2Q)+\sqrt{(b-2Q)^2+4(1-Q^2)}}{2(1-Q^2)}$. The equilibrium points $E_{1, 2, 3, 4}$ emerge at $\epsilon=\epsilon_{HB}$ through HB. The steady state $E_{NHSS}$ is stable if $-\frac{b}{2}<\epsilon<\frac{-(b-2Q)+\sqrt{(b-2Q)^2+4(1-Q^2)}}{2(1-Q^2)}$. Using eigenvalue analysis, the symmetry breaking coupling dependent steady state $E_{3, 4}$ is stable if $b+2 \epsilon>0, \epsilon(b+\epsilon)-k_2(1+\epsilon Q)>0 $ where $k_2={\epsilon Q}-2+3\epsilon(b+\epsilon)$. From eigenvalue analysis, we derive the Hopf bifurcation curve as $b+2\epsilon=0$ and inverse pitchfork bifurcation curve as $\epsilon^2(1-Q^2)+\epsilon(b-2Q)-1=0.$

 \begin{figure}[ht]
 	\centerline{
 		\includegraphics[scale=0.4]{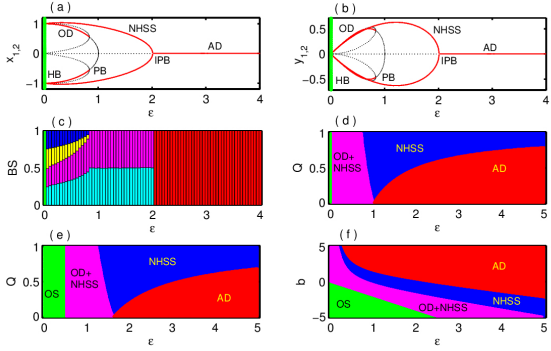}}
 	\caption{ Two Duffing-Holmes oscillators coupled through cross mean-field coupling: bifurcation diagrams by varying the cross mean-field coupling strength $\epsilon$ where extrema of (a)     $x_{1,2}$ and (b) $y_{1, 2}$ are plotted for $b=-0.01$ and $Q=0.5$. (c) Variation of BS for different values of $\epsilon$ where regions of color green represents oscillatory state, cyan and magenta for the steady states $E_{1, 2}$, blue and yellow for the steady states $E_{3, 4}$ and red for amplitude death state. Two parameter bifurcation diagrams in the $\epsilon-Q$ plane for (d) $b=-0.01$ and (e) $b=-1.0$. (f) Two parameter bifurcation diagram in the $\epsilon-b$ plane for $Q=0.5.$ The region of green, magenta, blue and red corresponding for the oscillatory, coexistence of OD and NHSS, solely NHSS and AD states respectively.}
 	\label{DH2_cross}
 \end{figure}     
 
 Figures~\ref{DH2_cross}(a) and (b) show the bifurcation diagrams with respect to coupling strength $\epsilon$ corresponding to the $x-$ and $y-$ components respectively with $b=-0.01$ and $Q=0.5$. As reflected in both of these figures, different coupling dependent fixed points, namely $E_{1,2}$ and $E_{3,4}$ resembling NHSS and OD states appear from oscillatory state through Hopf bifurcation at $\epsilon_{HB}$. Further increment in $\epsilon$ gives rise to an inverse pitchfork bifurcation at $\epsilon_{IPB}$ through which the saddle  point (the origin) gets stabilized that signifies the AD state. The variation in the BS of the steady states $E_{1,2}$ and $E_{3,4}$ with coupling strength $\epsilon$ are depicted in Fig.~\ref{DH2_cross}(c). For very small value of $\epsilon$, after the occurrence of Hopf bifurcation at $\epsilon_{HB}$, initially all the fixed points $E_{1,2}$ and $E_{3,4}$ share almost the same BS but as $\epsilon$ increases BS of $E_{3,4}$ starts decreasing and becomes zero at $\epsilon_{PB}$. From then the BS of $E_{1,2}$ remains same and further abruptly turns into zero at $\epsilon_{IPB}$. Further hike in $\epsilon$ leads the BS of  saddle point $E_0$ to be unity. $Q=0.5$ and $b=-0.01$ are kept fixed in this case.
 \par Figure~\ref{DH2_cross}(d) represents the parameter region in $\epsilon-Q$ plane in which green, magenta, blue and red regions respectively correspond to the  oscillatory state, coexistence of OD and NHSS, stable NHSS state and AD state for $b=-0.01$. As can be seen, here oscillation exists for all values of $Q$ but for a very narrow range of $\epsilon$ only. At the right of the Hopf bifurcation curve $b+2\epsilon=0$, firstly coupling dependent fixed points get stabilized and then AD state come out depending on the value of $Q$.   However, we choose the value of $b$ smaller than that used in Fig.~\ref{DH2_cross}(d), namely we take $b=-1.0$ and plot the $\epsilon-Q$ parameter plane in Fig.~\ref{DH2_cross}(e). As expected, because of the form the HB curve $b+2\epsilon=0$, a broader range of $\epsilon$ is now indicating oscillation of the coupled system in color green, in this case.  The other rengions of  coexistence of OD and NHSS, stable NHSS state and AD state are plotted in  magenta, blue and red colors respectively as before.
 \par Finally we plot the $\epsilon-b$ parameter plane in Fig.~\ref{DH2_cross}(f) while keeping $Q=0.5$ fixed. Only the negative values of $b$ can produce oscillation as shown in the figure. The process of stabilization of saddle point implying AD state through the stabilization of other coupling dependent fixed points indicating OD and NHSS state for almost all the values of $b$ is visible here.

 \section {Effect of Noise on steady states}
 In the main text, we analyze the stability  and probabilistic dominance of each steady states. We will analyze here the impact of noise in those steady states.
 Here we consider two DH oscillator coupled through mean-field coupling with additive common noise at each variable in the following form:
 \begin{equation}
 \begin{array}{lcl}
 \dot x_1=y_1+\epsilon(Q\frac{x_1+x_2}{2}-x_1)+\xi(t)\\
 \dot y_1=x_1-x_1^3-by_1+\epsilon(Q\frac{y_1+y_2}{2}-y_1)+\xi(t)\\
 \dot x_2=y_2+\epsilon(Q\frac{x_1+x_2}{2}-x_2)+\xi(t)\\
 \dot y_2=x_2-x_2^3-by_2+\epsilon(Q\frac{y_1+y_2}{2}-y_2)+\xi(t)
 \label{DH_noise}
 \end{array}.
 \end{equation}
 where $\epsilon$ is the mean-field coupling strength, $Q(0 \leq Q < 1)$ is the mean-field intensity, $\xi(t)$ is the Gaussian white noise with the properties $\langle \xi(t) \rangle = 0$ and $\langle \xi(t) \xi(t') \rangle = 2D \delta(t-t')$ where $D>0$ is the noise intensity, $\delta$ is Dirac delta function, and $\langle . \rangle$ denotes averaging over the realizations of $\xi(t)$.
 \par  At first we will prove why   stabilization of fixed points (linear stability approach) under noise is nearly impossible. However, numerically  we will show that  for a broader range of  noise interaction  the new states oscillate around the old states with a negligible fluctuations therefore the BS remains almost same as before.  \\
 
 {\bf Theorem : } {\it Oscillation suppression is impossible for noise induced continuous dynamical system. }
 \par
 {\bf {Proof :}} Consider ${\bf \dot {X}=f(X,\mu)}$ be $m-$dimensional continuous dynamical system, $\mu$ be it's system parameter. Let ${\bf X=X^*}$ be a stable equilibrium point for some  value of $\mu$. So the system will converge to the equilibrium point (${\bf X^*}$) for  any local perturbation of ${\bf X(t)}$ near the equilibrium point i.e. ${\bf \dot X(t)} = 0$ as $t \rightarrow \infty.$ The zero velocity of the system signifies the stabilization of  ${\bf X(t)}$ at that point.
 \par Now if we introduce additive noise in that system then it follows
 $${\bf \dot {X}=f(X,\mu)}+B{\bf\xi(t)},$$
 where $B$ is a diagonal matrix of order $m$, $\xi(t)$ be the noise function which is  time dependent.
 For an arbitrary initial condition, let's assume  the system vector ${\bf X(t)}$ arrives at ${\bf X(t_0)}$ when $t=t_0$. Then the velocity at the time $t=t_0$ becomes 
 $${\bf \dot {X}(t_0)=f(X(t_0), \mu)+B\xi(t_0)}.$$
 Now since ${\bf \xi(t)}$  fluctuates with time $t$ and it is independent of the evolution function ${\bf f(X, \mu)}$ so ${\bf \dot X}$ will never be constant for all subsequent time $t>t_0$. Particularly,  it will never converges to zero as time increases. The $non-zero$ velocity will force 
 the system to oscillate i.e. $\: {\bf X(t)}$ will never be stabilized as $t$ grows up.
 \par The above mathematical logic emphasizes that oscillation suppression is impossible for noise induced dynamical systems. $\square$

\begin{figure}
 	\includegraphics[width=85mm]{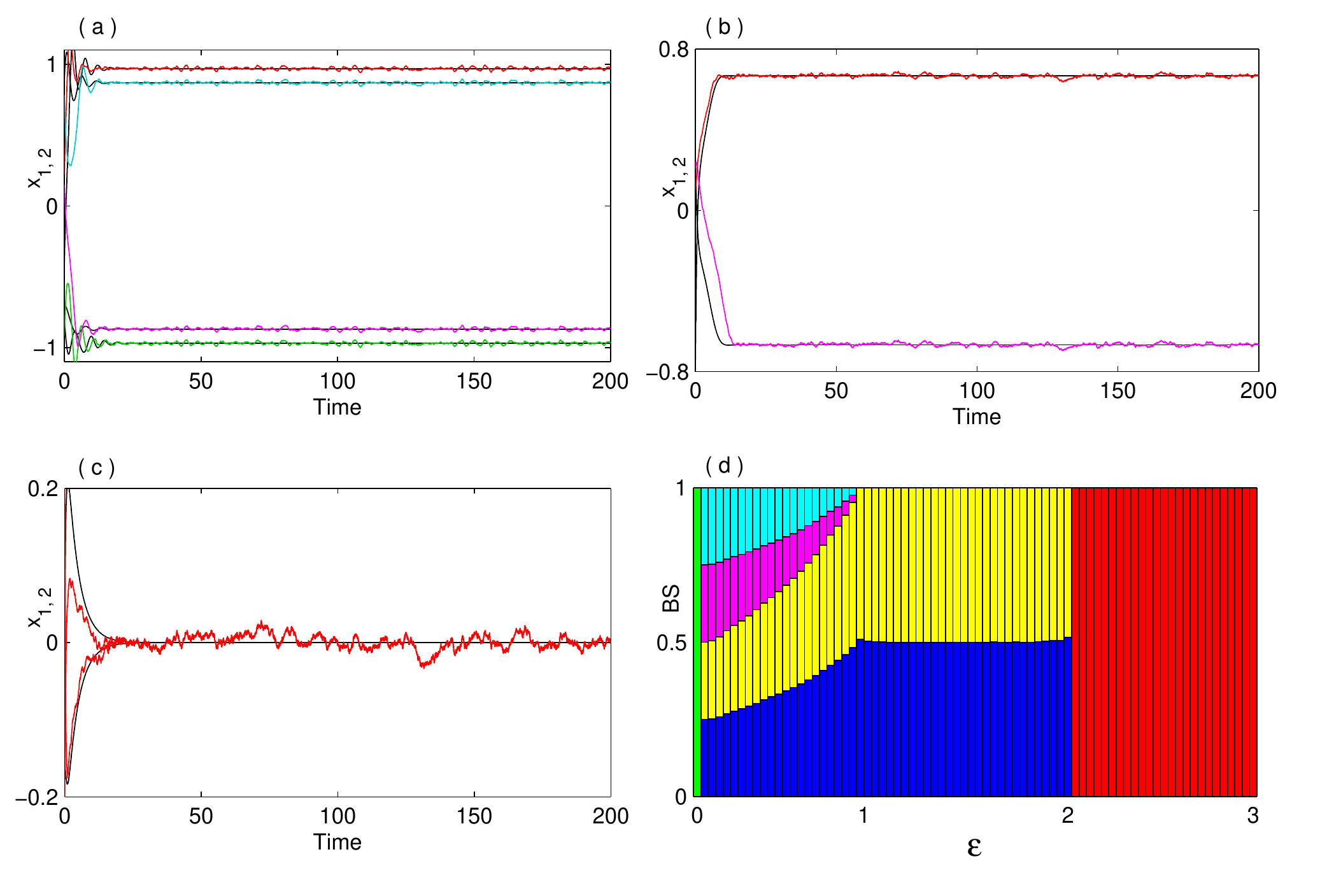}
 	\caption{Two coupled DH oscillator with common noise: Time series of the state variable $x_{1,2}$ for (a) $\epsilon = 0.5$, (b) $\epsilon = 1.5$, (c) $\epsilon = 2.5$ where black straight line shows the time series of steady states of noise-free ($D=0$) system, other color curves be the time series of the noise induced system for different initial conditions. (d) Variation of BS of these fluctuation state for various value of coupling strength $\epsilon$. The color region green, cyan, magenta, yellow, blue, red corresponds to oscillation state, fluctuations around the steady states $E_4,E_3,E_2,E_1$ and $E_0$. Other parameters: $b=-0.01, Q=0.5$, noise intensity $D=0.08$.}
 	\label{fig:fig1}
 \end{figure}

 We numerically check the effect of noise in DH oscillators (Eqn. \ref{DH_noise}).  Figure \ref{fig:fig1} describes the effect of noise on the steady states present in the systems.  The system parameters are  same (follow the main text). The noise intensity is taken as $D=0.08$. We check three coupling regimes ($\epsilon$) where the steady states are structurally different. In Fig. \ref{fig:fig1}(a) four states (two NHSS and two IHSS shown in black line) are plotted as a function of time (coupling strength $\epsilon=1.0$) and we observe   small oscillations (shown in four colors around the steady states (black lines)) due to the presence of noise.    Figure \ref{fig:fig1}(b)  reveals the nature of NHSS at $\epsilon=1.5$ where IHSS do not exist. The small oscillations exist around the steady states due to the presence of noise. Further Fig. \ref{fig:fig1}(c) explores the behavior of  one steady state:  the stabilization of the saddle equilibrium with noisy fluctuation.
 It seems for all cases the noise effect is statistically negligible as all the time series oscillate around the steady states with negligibly small amplitude.  We finally check how the Basin Stability  of these fluctuation states appear in the basin volume.  Figure 1(d) reveals the BS measure of such  fluctuations  around the steady states $E_{1,2}, E_{3,4}$ and $E_0$ as a function of $\epsilon$. The qualitative and quantitative behavior of this BS is exactly same with the BS scenario of the noise-free system. All the bifurcation points (HB, IPB) also appear in the same points.\\
 
 \begin{figure}
 	\includegraphics[width=90mm]{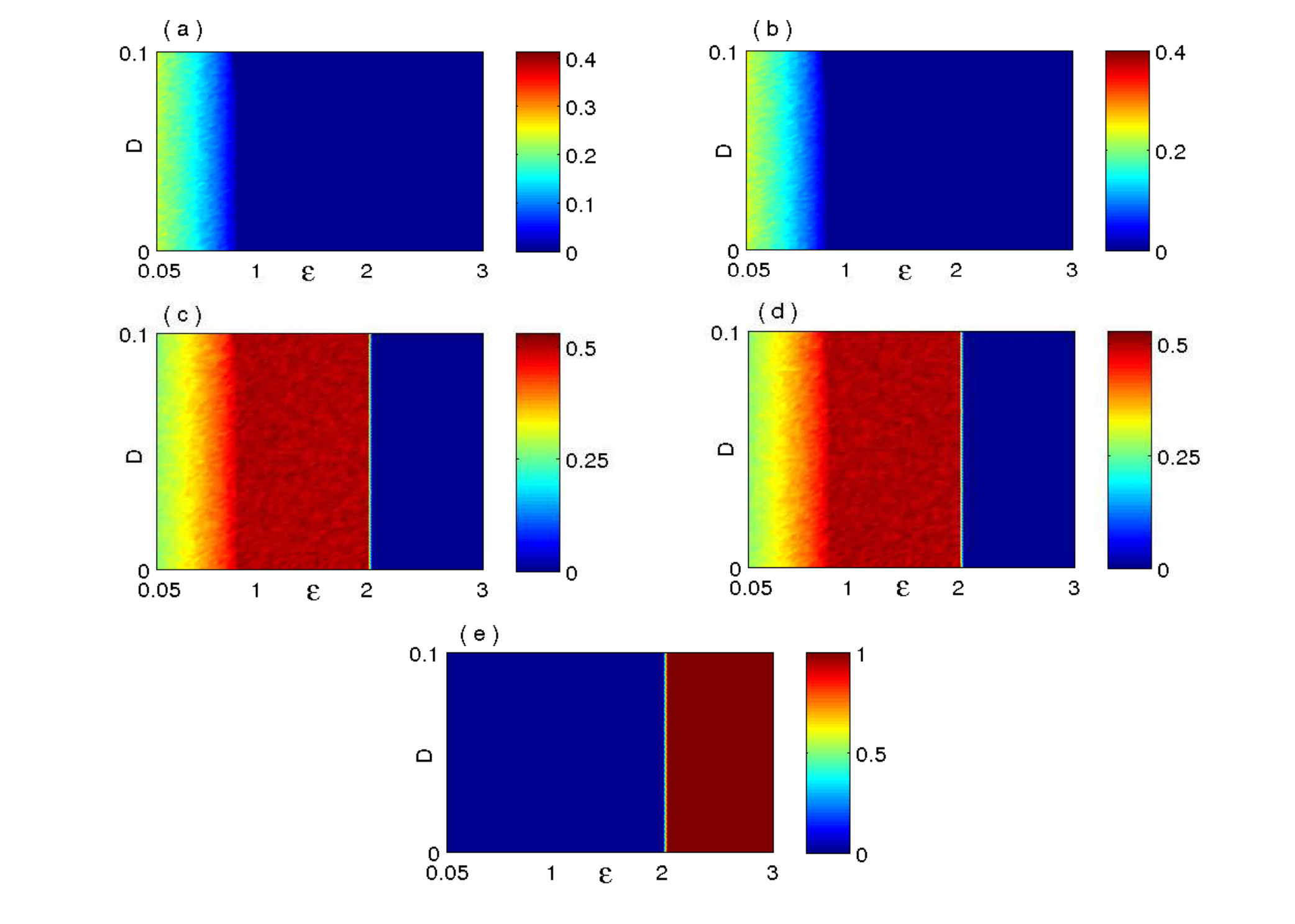}
 	\caption{Coexistence of different fluctuation states around fixed points are quantified by a color bar of basin stability measure in $\epsilon-D$ parameter space of two coupled DH oscillator with common noise; $b=-0.01, Q=0.5$: where (a), (b), (c), (d) and (e) represent the BS of fluctuated states around $E_4, E_3, E_2, E_1$ and $E_0$ respectively.}
 	\label{fig:fig2}
 \end{figure}
 \par 
 
 In the $\epsilon-D$ parameter space,   the BS of noise induced fluctuation states around $E_4, E_3, E_2, E_1$ and $E_0$ are shown in the color coded Figs.~\ref{fig:fig2}(a), (b), (c), (d) and (e) respectively. Figures~\ref{fig:fig2}(a), (b) show that just after the Hopf bifurcation point the BS of noise induced $E_{3,4}$  takes the value almost equal to $0.25$ for all $D\in[0,0.1]$.   Actually each of them acquires 25 percent of the whole space because four steady states ($E_{1,2,3,4}$) coexist together.
 However, the BS of these sates ($E_{3,4}$) gradually decrease  for more increasing  $\epsilon$, and finally they become unstable at $\epsilon=\epsilon_{PB}$. And  we observe that there is no impact of noise intensity on it's BS. The BS scenario of the fluctuation states around $E_{1,2}$ are shown in Figs.~\ref{fig:fig2}(c), (d).  After the Hopf bifurcation point they take value approximately $0.25$ for any noise intensity ($D\in[0,0.1]$).  Further increase   of $\epsilon$ the BS of these sates increase implying their more accessibility in the basin volume.  After $\epsilon=\epsilon_{PB}$  these two states becomes bi-stable as $E_{3,4}$ lose their stability irrespective of the presence of noise  $D\in[0,0.1]$. From  Fig.~\ref{fig:fig2}(e) we can see that the BS of fluctuation  state around $E_0$ is equal to 0 before the coupling strength less than $\epsilon_{IPB}$, but it abruptly becomes stable at $\epsilon=\epsilon_{IPB}$ with BS equal to 1. But there is no change of BS along the $Y-$axis which signifies that no significant influence of noise on it's BS.


\end{document}